\documentclass[prb,aps,showpacs,twocolumn,floatfix]{revtex4-1}
\usepackage{amsmath,amsfonts,amssymb,graphicx,bm}
\newcommand{\Psid}{\Psi^\dag}
\newcommand{\e}{{\rm e}}
\newcommand{\ch}{{\rm ch}}
\newcommand{\sh}{{\rm sh}}

\newcommand{\vkp}{{\bm k}_\perp}
\newcommand{\ssp}{s_\perp}
\newcommand{\ep}{\epsilon_\perp}
\newcommand{\ri}{{\bm x}_i}
\newcommand{\z}{\mathbb{Z}_2}
\begin{document}
\title{Bulk-boundary correspondence in three dimensional topological
       insulators}
\author{L. Isaev}
\author{Y. H. Moon}
\author{G. Ortiz}
\affiliation{Department of Physics, Indiana University, Bloomington IN 47405}

\begin{abstract}
 We discuss the relation between bulk topological invariants and the spectrum
 of surface states in three dimensional non-interacting topological insulators.
 By studying particular models, and considering general boundary conditions for
 the electron wavefunction on the crystal surface, we demonstrate that using
 experimental techniques that probe surface states, only strong topological and
 trivial insulating phases can be distinguished; the latter state being
 equivalent to a weak topological insulator. In a strong topological insulator,
 only the {\it parity} of the number of surface states, but not the number
 itself, is robust against time-reversal invariant boundary perturbations. Our
 results suggest a $\z$ definition of the bulk-boundary correspondence,
 compatible with the $\z$ classification of topological insulators.
\end{abstract}
\pacs{73.20.-r, 73.21.Fg, 73.22.Dj}
\maketitle

\section{Introduction}
\label{sec:introduction}
The defining characteristic of an electric insulator is the existence of an
energy gap to charge excitations. Depending on the physical origin of that gap
insulating materials are broadly divided into two classes: Mott insulators
with the gap having an origin in the electron interactions, and band
insulators, where the gap originates essentially from the single particle
energy terms, with many-body effects simply renormalizing the bare band
parameters. Because of their single-particle nature band insulators are often
thought to be the simplest systems, whose electronic properties are adequately
described by the usual quantum theory of solids \cite{Landau_IX}. Recently,
however, a classification of these materials has emerged
\cite{Kane_colloquium,Moore_3DTI}, based on topological invariants
\cite{Thouless_1998}, which characterize their band structure. In particular,
it was shown \cite{Kane_Z2,Zhang_MEE} that spin-orbit (SO) interaction and
time-reversal symmetry can stabilize topologically non-trivial electronic
states in certain systems, which were thus termed topological insulators (TI).
An experimental signature of these phases is the presence of chiral metallic
surface states, which are claimed to be robust against time-reversal invariant
local perturbations and weak disorder. Surface states were indeed observed in
bismuth antimony alloys \cite{Moore_3DTI}, as well as in the ${\rm Bi_2X_3}$
and ${\rm Sb_2X_3}$, with ${\rm X=(Se,Te)}$, families
\cite{Zhang_2009,Hsieh_2009}.

The bulk-boundary correspondence is a physical concept which relates
topological properties of the bulk with the number of gapless edge modes. This
relation has generated a lot of discussion in the context of Quantum Hall
Effect \cite{Halperin_1982,Hatsugai_1993}, graphene \cite{Hatsugai_2007}, and
now TI \cite{Kane_Z2,Zhang_MEE,Zhang_2006,Mong_2011}. In particular, in Ref.
\onlinecite{Kane_Z2} it was argued that a two-dimensional TI is characterized
by an odd number of metallic edge states. Qualitatively, due to time-reversal
symmetry, surface states in STI do not back-scatter and thus, can not be easily
localized, becoming robust against local time-reversal invariant perturbations.
In three dimensions (3D) it is argued that one should distinguish
\cite{Kane_Z2} among strong and weak TI (STI and WTI) states, where the Fermi
arc encloses an odd or even number of surface Dirac (i.e. Kramers-degenerate)
points, respectively.

So far the bulk-boundary correspondence remains a quite loosely formulated
conjecture. It remains unclear how universally valid is this conjecture in real
many-body insulating systems, or how robust is this correspondence with respect
to variations in the surface properties of those systems. In any bulk-boundary
correspondence specific properties of the surface should be of relevance.
Indeed, the electronic edge spectrum can be very sensitive \cite{Volkov_1981}
to the specific form of the effective surface Hamiltonian. For example, in the
context of graphene the connection between the valley-specific Hall
conductivity and the number of gapless edge modes has been found to be
dependent upon the boundary conditions (BC) \cite{Martin_2010}.

In the present paper we examine the bulk-boundary correspondence in 3D band
insulators with strong SO interaction. We present a technique to analyze their
surface spectra based on the ideas of Ref. \onlinecite{Satanin_1984}, developed
in the context of semiconductor nanostructures, and study the effect of
variations in BC on the spectrum of surface states. In particular, we use
algebraic properties of the model Hamiltonian to construct BC according to
symmetries of the problem, such as time-reversal invariance. We also discuss
the role of {\it symmetry-breaking} surface perturbations. Our work leads to a
two-fold result. On the one hand, we show that even in a clean system with no
surface reconstruction, from the standpoint of the surface spectra, there
exists no physical distinction between WTI and topologically trivial insulating
phases. In particular, we present examples of trivial insulators, which would
appear as WTIs, since they also possess an even number of surface Dirac cones.
On the other hand, the STI state is robust against time-reversal invariant
boundary perturbations, in the sense that the Fermi arc always encloses an odd
number of surface Dirac points with an odd number of edge states crossing the
Fermi level along a path between two time-reversal invariant momenta (TRIMs) in
the surface Brillouin zone. The number of crossings depends on the particular
choice of BC. This indicates that edge states in a STI are {\it not robust}
against arbitrary time-reversal invariant surface perturbations. However, the
{\it parity} of their number is protected.

These observations provide a $\z$ formulation of the bulk-boundary
correspondence in 3D TI:
\begin{displaymath}
 \nu_0=N_s\,\,{\rm mod}\,\,2,
\end{displaymath}
where $\nu_0$ is the strong $\z$ bulk topological invariant \cite{Kane_Z2},
$N_s$ is the number of Kramers-degenerate points in the surface Brillouin zone
enclosed by the Fermi arc, and we assume that there are no time-reversal
breaking perturbations at the insulator's surface. This relation shows that it
is only appropriate to differentiate between STI and trivial insulating phases,
characterized by $\nu_0=1$ and $\nu_0=0$, respectively. From an experimental
point of view (by looking at the spectrum of surface states), a WTI is
equivalent to a trivial insulator as will be shown.

The paper is organized as follows. In the next section we consider a simple
example of a trivial insulator and show that under certain conditions it may
possess an even number of metallic edge states, thus appearing as a WTI. In
Sec. \ref{sec:franz_model} we discuss the robustness of surface states in a
model of a STI, and demonstrate the {\it protection} of parity of their number,
but not the number or the states themselves. Moreover, we further discuss the
indistinguishability between a WTI and a trivial insulator from the standpoint
of the number of edge (surface) modes. In Sec. \ref{sec:tri_break}, we study
the effect of time-reversal breaking surface perturbations on the edge spectrum
of a STI. Our conclusions are summarized in Sec. \ref{sec:conclusion}.

\section{Dirac-like band model}
\label{sec:dirac_model}
The SO interaction plays an important role in establishing a band structure
characterized by non-trivial topological invariants. Physically, this means
that proper interband matrix elements of the SO term in the Hamiltonian must be
comparable with the bulk band gap, so that the bands become essentially
non-parabolic. As a ubiquitous consequence of the band mixing, Dirac-like bulk
spectra of electrons and holes are formed \cite{Bir_Pikus}. In this section we
consider a simple model of a topologically trivial insulator, which neverthless
would look like a WTI in the sense that the Fermi arc encloses an even number
of surface Dirac points. We also show that the existence of edge states is very
sensitive to the BC, imposed on the single-particle wavefunction at the crystal
surface.

\subsection{General formalism}
The simplest lattice model, which contains SO interaction naturally, is just
the tight-binding form of the Dirac Hamiltonian:
\begin{align}
 H_D=\frac{\lambda}{2{\rm i}}\sum_{i,\mu}\bigl(\Psid_i\alpha^\mu\Psi_{i+\mu}
 \e^{{\rm i}\theta^\mu_i}-{\rm h.c.}\bigr)+\epsilon\sum_i\Psid_i\beta\Psi_i,
 \label{dirac_hamiltonian}
\end{align}
where from now on subindex $i\equiv{\bm x}_i=(x_i,y_i,z_i)$ enumerates sites of
a simple cubic lattice; $i+\mu\equiv\mu_i+1$ with $\mu=x,y,z$;
$\Psi_i=(c_{ic\sigma},c_{iv\sigma})$ destroys an electron with spin $\sigma$ in
the conduction ($c$) or valence ($v$) band; $\lambda$ and $\epsilon$ are the SO
coupling constant and band gap, respectively; and $\theta^\mu_i$ is a $U(1)$
gauge field. The Dirac matrices $\beta$ and $\alpha^\mu$ have the well-known
properties \cite{Akhiezer_Berestetskii}:
\begin{align}
 \{\alpha^\mu,\alpha^\nu\}=2\delta_{\mu\nu};&\,\,\{\alpha^\mu,\beta\}=0;\,\,
 [\alpha^\mu,\alpha^\nu]=2{\rm i}\varepsilon_{\mu\nu\kappa}\Sigma_\kappa;
 \nonumber \\
 &(\alpha^\mu)^2=\beta^2=1 \nonumber
\end{align}
with $\Sigma_\mu=(1\otimes\sigma^\mu)$ being the 4-spin operator, $\sigma^\mu$
-- the Pauli matrix, and $\varepsilon_{\mu\nu\kappa}$ -- the L\'evi-Civita
symbol. From Eq. \eqref{dirac_hamiltonian} it is easy to obtain an expression
for the charge current, as a variation of the Hamiltonian with respect to the
gauge field:
\begin{equation}
 J^\mu_i\equiv\frac{\delta H_D}{\lambda\delta\theta^\mu_i}\bigg
 \vert_{\theta^\mu_i=0}=\frac{1}{2}\bigl(\Psid_{i}\alpha^\mu\Psi_{i+\mu}+
 \Psid_{i+\mu}\alpha^\mu\Psi_i\bigr).
 \label{dirac_current}
\end{equation}
In the rest of this section, we assume that $\theta_i^\mu=0$.

The Hamiltonian of Eq. \eqref{dirac_hamiltonian} commutes with the
time-reversal (${\cal T}$) and space inversion (${\cal P}$) operators. Also,
the bulk band structure is invariant under interchange of the conduction and
valence bands, which means that $H_D$ has the charge-conjugation (${\cal C}$)
symmetry. These three operations are defined by their action on a
single-particle orbital $\psi(t,\ri)=\psi_i(t)$ \cite{Akhiezer_Berestetskii}:
\begin{align}
 {\cal T}\psi(t,\ri)=\alpha^x&\alpha^z\psi^*(-t,\ri);\,\,
 {\cal P}\psi(t,\ri)=\beta\psi(t,-\ri); \nonumber \\
 &{\cal C}\psi(t,\ri)=\beta\alpha^y\psi^*(t,\ri). \nonumber
\end{align}
Motion in an infinite crystal also preserves the tensor spin operator
\cite{Satanin_1984,Bagrov_Gitman}:
\begin{equation}
 T_\mu=\frac{\varepsilon_{\mu\nu\kappa}}{2{\rm i}}\sum_i\bigl(\Psid_i\beta
 \Sigma_\nu\Psi_{i+\kappa}-{\rm h.c.}\bigr).
 \label{T}
\end{equation}
It follows that ${\bm T}$ is a polar, time-reversal invariant vector. 

\begin{figure}[t]
 \begin{center}
  \includegraphics[width=\columnwidth]{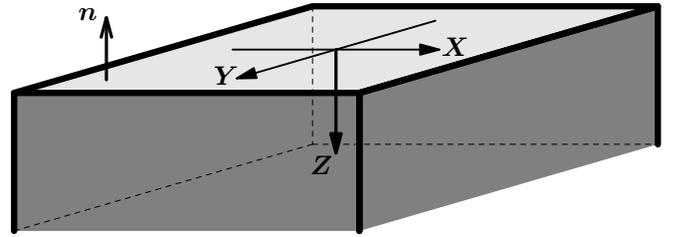}
 \end{center}
 \caption{Geometry of the problem studied in this paper. The crystal is bounded
          by the plane $z_i=0$ and occupies the half-space $z_i>0$. The unit
          vector ${\bm n}$ denotes an outer normal to the surface.}
 \label{fig_geometry}
\end{figure}

In order to study the spectrum of surface states, we now consider a
half-infinite crystal, bounded by the plane $z_i=0$ and occupying the region
$z_i>0$ (see Fig. \ref{fig_geometry}). On the surface the single-particle
wavefunction $\psi_i$ will satisfy a linear relation of the form:
\begin{equation}
 -\sum_\mu n_\mu\psi_{i_s+\mu}+S\psi_{i_s}=0,
 \label{dirac_bc}
\end{equation}
where ${\bm n}$ is the outer normal to the surface and $i_s$ denotes 
lattice sites on the surface. The general structure of the $4\times4$ boundary 
matrix $S$ is fixed by requirements \cite{Volkov_1981,Satanin_1984} that the
current $\langle {\bm J}_0\cdot{\bm n}\rangle$ through the surface vanishes,
and the fundamental symmetries ${\cal T}$ and ${\cal P}$ are preserved:
\begin{equation}
 S=\xi_0\beta+{\rm i}\xi_1\beta{\bm\alpha}\cdot{\bm n}.
 \label{dirac_S}
\end{equation}
In this expression the parameters $\xi_{0,1}$ are free. At the phenomenological
level, they encode various surface properties. For example, the first term in
Eq. \eqref{dirac_S} changes sign under charge conjugation ${\cal C}$, which
physically means that at the boundary there is a mixing (whose amount is
controlled by $\xi_0$) of the bulk Bloch bands. Similarly, the second term in
$S$ describes the intensity of SO interaction at the surface, which is caused
by rapid changes in the crystal field. Thus, after scattering from the surface
an electron acquires an extra phase, due to spin rotation. Since the localized
(Tamm) states at the crystal boundary are formed as a result of interference of
bulk Bloch states \cite{Tamm_1932}, this term has a profound effect on their
stability. We also note that the BC \eqref{dirac_S} is translationally
invariant along the surface.

In the bounded crystal in Fig. \ref{fig_geometry}, only the normal to the
surface component of ${\bm T}$ [Eq. \eqref{T}] is conserved. In ${\bm k}$-space
(${\bm k}$ is the crystal momentum) it can be written as:
\begin{displaymath}
 T_z=\sum_{\vkp,z_i}\Psid_{z_i}(\vkp)t^z(\vkp)\Psi_{z_i}(\vkp)
\end{displaymath}
with $\vkp=(k_x,k_y)$, and
\begin{displaymath}
 t^z(\vkp)=\beta\bigl(\Sigma_x\sin k_y-\Sigma_y\sin k_x\bigr).
\end{displaymath}
The eigenstates of $t^z(\vkp)$ have the form:
\begin{equation}
 \psi_{\tau\vkp}(z_i)=
 \begin{pmatrix}
  \varphi_{\tau\vkp}(z_i)U_{\tau\vkp} \\
  \chi_{\tau\vkp}(z_i)U_{-\tau\vkp}
 \end{pmatrix},
 \label{tz_eigenstates}
\end{equation}
where
\begin{displaymath}
 U_{\tau\vkp}=\frac{1}{\sqrt{2}}
 \begin{pmatrix}
  1 \\
  -\frac{{\rm i}\tau}{\ssp}(\sin k_x+{\rm i}\sin k_y)
 \end{pmatrix},
\end{displaymath}
$\tau=\pm1$, $\ssp=\sqrt{\sin^2k_x+\sin^2k_y}$, and the amplitudes $\varphi$
and $\chi$ are arbitrary. The corresponding eigenvalues are $\tau\ssp$. The
$z$-independent part of the kinetic energy in Eq. \eqref{dirac_hamiltonian} can
be expressed in terms of $t^z(\vkp)$:
\begin{align}
 \frac{\lambda}{2{\rm i}}\sum_{i,\mu=(x,y)}\bigl(\Psid_i\alpha^\mu&\Psi_{i+\mu}
 -{\rm h.c.}\bigr)= \nonumber \\
 &={\rm i}\lambda\sum_{\vkp,z_i}\Psid_{\vkp z_i}\beta\alpha^zt^z(\vkp)
 \Psi_{\vkp z_i}.
 \nonumber
\end{align}
This fact has two important consequences: (i) The problem becomes
{\it effectively one-dimensional} and we can work with two-component
wavefunctions, similar to the case of a spherically symmetric field
\cite{Akhiezer_Berestetskii}:
\begin{displaymath}
 \begin{pmatrix}
  \varphi_{\tau\vkp}(z_i)U_{\tau\vkp} \\
  \chi_{\tau\vkp}(z_i)U_{-\tau\vkp}
 \end{pmatrix}
 \to
 \begin{pmatrix}
  \varphi_{\tau\vkp}(z_i) \\
  \chi_{\tau\vkp}(z_i)
 \end{pmatrix}.
\end{displaymath}
(ii) We are free to choose a representation for the Dirac matrices $\beta$
and $\alpha^z$. In the rest of the section, we use the following
identification:
\begin{displaymath}
 \beta\to\sigma^z;\quad\alpha^z\to\sigma^x.
\end{displaymath}

\begin{figure}[t]
 \begin{center}
  \includegraphics[width=\columnwidth]{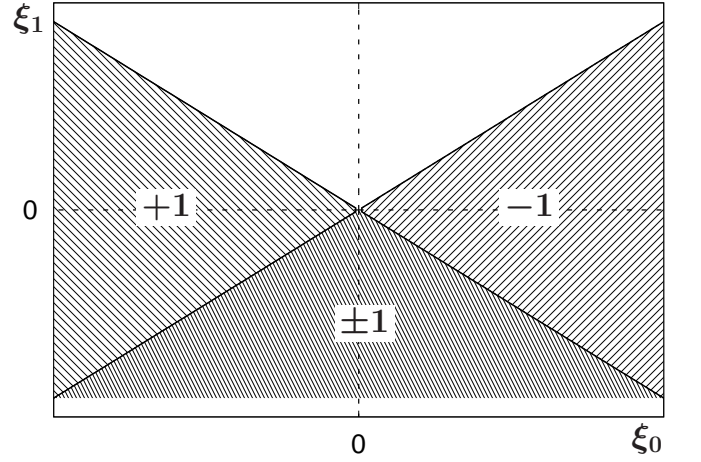}
 \end{center}
 \caption{Stability regions of edge states, obtained from Eq.
          \eqref{dirac_ss_spectrum}. Inside shaded areas there exist surface
          states with at least one value of the polarization $\tau$, as
          indicated by the numbers $\pm1$. The boundaries of shaded regions are
          given by the expression $\xi_1=-(\lambda\tau\ssp/\epsilon)\xi_0$.}
 \label{fig_dirac_ss_stability}
\end{figure}

\subsection{Surface spectrum}
In order to obtain the edge (surface) spectrum, we follow the method of Ref.
\onlinecite{Tamm_1932}. First, let us consider the bulk band structure:
\begin{displaymath}
 \varepsilon_{\vkp,k_z}=\pm\sqrt{\epsilon^2+\lambda^2(\ssp^2+\sin^2k_z)}.
\end{displaymath}
A surface state will have a complex-valued $k_z=p+{\rm i}q$ with $p$ and $q$
chosen in such a way that ${\rm Im}\,\varepsilon=0$. Since $q\neq0$, one
obtains: (i) $p=0$ or $p=\pi$, and (ii) $p=\pm\pi/2$. Only in case (i) the
state has an energy inside the band gap. Therefore, a general (two-component)
wavefunction, decaying into the crystal ($z_i>0$), can now be written as a
linear combination:
\begin{displaymath}
 \psi_{\tau\vkp}(z_i)=\sum_{p=0,\pi}c_p
 \begin{pmatrix}
  \varepsilon+\epsilon \\
  {\rm i}\lambda(\e^{{\rm i}p}\sh\,q-\tau\ssp)
 \end{pmatrix}
 \biggr)\e^{({\rm i}p-q)z_i}.
\end{displaymath}
Imposing the BC of Eqs. \eqref{dirac_bc} and \eqref{dirac_S} one easily obtains
the inverse localization length $q$ of the state (cf. Ref.
\onlinecite{Volkov_1981}):
\begin{equation}
 \sh^2q_{\tau\vkp}=\frac{r\varrho_\tau-1/2+\sqrt{(\varrho_\tau-r/2)^2+
 (1-r^2)/4}}{1-r^2},
 \label{dirac_ss_spectrum}
\end{equation}
which defines the surface state dispersion relation via
$\varepsilon_{\tau\vkp}=\pm\sqrt{\epsilon^2+
\lambda^2(\ssp^2-\sh^2q_{\tau\vkp})}$.
In this expression:
\begin{align}
 \varrho_\tau(\xi_0,\xi_1)=&(1-r)(\tau\ssp\xi_0+\epsilon\xi_1/\lambda);
 \nonumber \\
 r(\xi_0,\xi_1)=&\frac{\xi_0^2+\xi_1^2-1}{\xi_0^2+\xi_1^2+1}\in(-1,1).\nonumber
\end{align}

A surface state exists only if $q>0$, which is equivalent to $\varrho_\tau<0$.
Regions in the plane $(\xi_0,\xi_1)$, where this inequality is satisfied, are
shown in Fig. \ref{fig_dirac_ss_stability}. For a given point inside a shaded
area, there is a solution either with one, or both values of $\tau$. Moreover,
there is an unshaded region, where the edge spectrum disappears. In general,
the surface states are gapped. However, this gap closes at some points in Fig.
\ref{fig_dirac_ss_stability}. For example, let us consider the case $\xi_0=0$,
$\xi_1<0$ and $r=-\eta/\sqrt{1+\eta^2}$ with $\eta=\epsilon/\lambda>0$. Then,
solutions for $p=0$ and $\pi$ disentagle. Therefore, $\sh\,q_{\tau\vkp}=\eta$,
$\varepsilon^p_{\tau\vkp}=\lambda\e^{{\rm i}p}\tau\ssp=\pm\lambda\tau\ssp$, and
the wavefunction is:
\begin{equation}
 \psi_{\tau\vkp}^p(z_i)=\sqrt{\eta}
 \begin{pmatrix}
  1 \\
  {\rm i}\e^{{\rm i}p}
 \end{pmatrix}
 \bigl[-\eta+\sqrt{1+\eta^2}\bigr]^{z_i+\frac{1}{2}}\e^{{\rm i}pz_i}.
 \nonumber
\end{equation}

Since $\varepsilon_{\tau\vkp}$ depends on $\vkp$ only through $\ssp$, for the
chemical potential inside the band gap, the Fermi arc always encloses an even
number of Dirac points. Thus, from an experimental perspective this model
system would look like a WTI \cite{Kane_Z2}. Nevertheless, it is
straightforward to check that {\it all four topological invariants}
$(\nu_0(\nu_1\nu_2\nu_3))$ {\it vanish} (we remind the reader that $\nu_0=0$
and $(\nu_1\nu_2\nu_3)\neq 0$ is the mathematical characterization of a WTI).
As we shall see in the next section, the above result is not specific to the
model of Eq. \eqref{dirac_hamiltonian}.

\section{Lattice Dimmock model}
\label{sec:franz_model}
The Dirac model \eqref{dirac_hamiltonian} describes a trivial insulator.
However, it can be extended to support a STI phase. We consider one such
modification, proposed in Ref. \onlinecite{Rosenberg_2010}:
\begin{equation}
 H=H_D-t\sum_{i,\mu}\bigl(\Psid_i\beta\Psi_{i+\mu}\e^{{\rm i}\theta^\mu_i}+
 {\rm h.c.}\bigr),
 \label{dimmock_hamiltonian}
\end{equation}
where $t>0$ is the hopping amplitude between nearest neighbor sites in a 3D
cubic lattice. The model thus defined is a lattice analog of the well-known
Dimmock model \cite{Dimmock_1964}, which provides an effective-mass description
of electronic spectra in lead chalcogenides. The Dirac model, studied in the
previous section, is recovered in the limit $t\to0$. However, the point
$t=0$ is singular and the limit has to be taken carefully.

Depending on the ratio of $\epsilon/t$, the model \eqref{dimmock_hamiltonian}
exhibits the following phases \cite{Rosenberg_2010}: (i) STI if
$2t<\vert\epsilon\vert<6t$, (ii) WTI for $\vert\epsilon\vert<2t$ and (iii) the
trivial band insulator when $\vert\epsilon\vert>6t$. In the STI phase the Fermi
energy inside the gap must cross an odd number of edge states along a path
between two time-reversal invariant momenta in the surface Brillouin zone.
Indeed, it was shown a long time ago that in the long wavelength approximation
under band inversion the Dimmock model has exactly one Dirac cone at the
surface \cite{Kisin_1987}. This conclusion is in agreement with the phase
diagram, because in the continuum limit the band gap $\Delta=\epsilon-6t$ is
negative for $\epsilon\lesssim6t$. In this section we study the effect of
variations in the BC on the surface spectrum of the lattice Dimmock model, and
on the physical meaning of the above phase diagram.

As before, we start by deriving the BC with given symmetry properties. 
The probability current
\begin{align}
 J^\mu_i=&\frac{1}{2}\bigl(\Psid_{i}\alpha^\mu\Psi_{i+\mu}+\Psid_{i+\mu}
 \alpha^\mu\Psi_i\bigr) \nonumber \\
 &-\frac{{\rm i}t}{\lambda}\bigl(\Psid_i\beta\Psi_{i+\mu}-
 \Psid_{i+\mu}\beta\Psi_i\bigr).
 \label{dimmock_current}
\end{align}
vanishes at the surface if the wavefunction satisfies the constraint
\eqref{dirac_bc} with a boundary operator of the form:
\begin{equation}
 S=\frac{2t}{\lambda}\xi_0+\xi_1\beta+{\rm i}\xi_0\beta{\bm\alpha}\cdot{\bm n}.
 \label{dimmock_S}
\end{equation}
The discussion presented after Eq. \eqref{dirac_S} regarding the physical
meaning of individual terms is fully applicable here as well.

It is easy to see that the lattice Dimmock model \eqref{dimmock_hamiltonian}
has the same symmetries as the Dirac model of the previous section. Therefore,
the problem of determining the surface spectrum, in the geometry of Fig.
\ref{fig_geometry}, again becomes one-dimensional. However, now it is
convenient to choose a different representation of the Dirac matrices:
\begin{equation}
 \alpha^z\to\sigma^y;\quad\beta\to\sigma^x.
 \label{dimmock_representation}
\end{equation}
We also introduce a notation $\ep=\epsilon-2t(\cos k_x+\cos k_y)$.

The bulk band structure has the form:
\begin{displaymath}
 \varepsilon_{\vkp,k_z}=\pm\sqrt{(\ep-2t\cos k_z)^2+
 \lambda^2(\ssp^2+\sin^2k_z)}.
\end{displaymath}
The requirement ${\rm Im}\,\varepsilon=0$ is equivalent to the equation
${\rm Im}\,\varepsilon^2=0$ under the condition
${\rm Re}\,\varepsilon^2\geqslant0$. In the complex $k_z$-plane
($k_z=p+{\rm i}q$) ${\rm Im}\,\varepsilon^2$ vanishes when: (i) $p=0$, (ii)
$p=\pi$ and (iii) along the line
$\cos p\,\ch\,q=2t\ep/\bigl[(2t)^2-\lambda^2\bigr]$. In general, a surface
state wavefunction is a complicated function with a $\vkp$-dependent complex
localization length. However, there exists a parameter range, where the edge
states are purely evanescent, i.e. $p=0$. Indeed, let us assume that
\begin{equation}
 4t<\epsilon<10t;\quad
 (2t)^2-\frac{1}{3}(\epsilon-4t)^2\leqslant\lambda^2\leqslant(2t)^2.
 \label{dimmock_parameters}
\end{equation}
Then, possible values of $q$ are given by:
\begin{align}
 \ch\,q_{1,2}=&\frac{2t\ep}{(2t)^2-\lambda^2} \nonumber \\
 &\mp\frac{\sqrt{\ep^2\lambda^2+\bigl[\varepsilon^2-\lambda^2(\ssp^2+1)\bigr]
 \bigl[(2t)^2-\lambda^2\bigr]}}{(2t)^2-\lambda^2}. \nonumber
\end{align}
We note that because of peculiar properties of the band structure, in the
present model there always exists a partial evanescent solution with
$\ch\,q_2>1$. However, by virtue of Eq. \eqref{dimmock_parameters} both
$\ch\,q_{1,2}\geqslant1$, with $q_1$ being the inverse localization length of a
state. Using the form \eqref{dimmock_representation} of $\alpha^z$ and $\beta$,
the general localized solution can be written as:
\begin{displaymath}
 \psi_{\tau\vkp}(z_i)=\sum_{a=1,2}c_a
 \begin{pmatrix}
  \ep-2t\ch\,q_a+\lambda\sh\,q_a \\
  \varepsilon+\lambda\tau\ssp
 \end{pmatrix}
 \e^{-q_az_i}.
\end{displaymath}

Before considering generic BC \eqref{dimmock_S}, let us analyze the case
$\xi_1=0$, when there is no band mixing at the boundary. One can easily show
that there exists a surface state with energy
$\varepsilon_{\tau\vkp}=\lambda\tau\ssp$, and momentum $q$
\begin{equation}
 \ch\,q_{1,2}=\frac{2t\ep}{(2t)^2-\lambda^2}\mp\frac{\sqrt{\lambda^2\ep^2-
 \lambda^2\bigl[(2t)^2-\lambda^2\bigr]}}{(2t)^2-\lambda^2},
 \label{dimmock_ss_spectrum}
\end{equation}
with the resulting wavefunction
\begin{equation}
 \psi_{\tau\vkp}(z_i)\sim
 \begin{pmatrix}
  0 \\
  1
 \end{pmatrix}
 \bigl(\e^{-q_1z_i}-\eta(\xi_0)\e^{-q_2z_i}\bigr),
 \label{dimmock_ss_state}
\end{equation}
where $\eta(\xi_0)=\bigl[\bigl(2t/\lambda-1\bigr)\xi_0+\e^{-q_1}\bigr]/\bigl[
\bigl(2t/\lambda-1\bigr)\xi_0+\e^{-q_2}\bigr]$. This state exists only in a
region of $\vkp$-space, defined by $\ep\leqslant 2t$. At the boundary of this
region $q_1=0$, and the state merges into the bulk continuum. Clearly, the
above condition is {\it a priori} false when $\epsilon>6t$, in agreement with
the general phase diagram \cite{Rosenberg_2010}.

\begin{figure}[t]
 \begin{center}
  \includegraphics[width=\columnwidth]{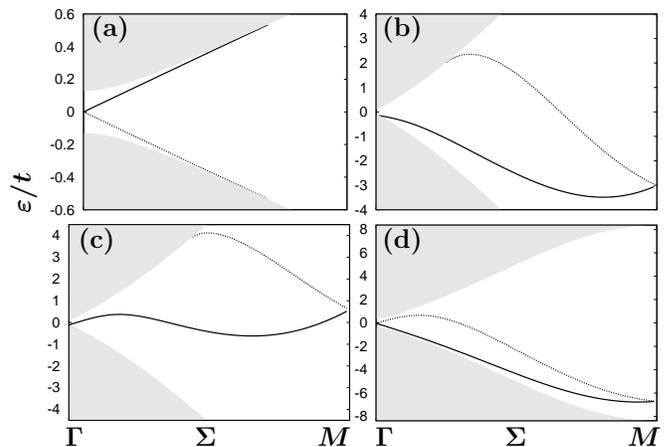}
 \end{center}
 \caption{Surface spectra of the lattice Dimmock model for different values of
          the parameters $\xi_{0,1}$ and $\epsilon/t$. SO interaction is
          $\lambda/t=\sqrt{3}$. The spectra are plotted along the path
          $(0,0)$-$(k_x,k_y)$-$(\pi,\pi)$ in the surface Brillouin zone. The
          shaded regions describe the bulk continuum with the band gap equal to
          $\vert\epsilon-6t\vert$. In panel (a) the plot is actually truncated
          before the $M$-point is reached, for readability. Solid lines
          represent $\tau=+1$, dashed lines $\tau=-1$. (a) $\xi_0=1.0$,
          $\xi_1=0.0$; (b) $\xi_0=0.0$, $\xi_1=0.1$; (c) $\xi_0=0.0$,
          $\xi_1=0.045$; (d) $\xi_0=0.1$, $\xi_1=0.3$. Panels (a)--(c)
          correspond to a STI with $\epsilon/t=5.87$, panel (d) describes a
          trivial insulator with $\epsilon/t=6.37$.}
 \label{fig_dimmock_sti_ss}
\end{figure}

When $\xi_{0,1}\neq0$ we observe that matrices $1$, $\beta$ and
${\rm i}\beta{\bm\alpha}\cdot{\bm n}$ consitute an orthonormal (though not
closed) set $\{m_i\}$: ${\rm Tr}\,m_im_j=4\delta_{ij}$, $m_i^2=1$. Thus, for
any value of $\xi_1$ there exists a unitary transformation
${\cal U}(\omega)=\exp{\bigl({\rm i}\omega{\bm\alpha}\cdot{\bm n}/2\bigr)}$,
which diagonalizes $S$ (cf. the discussion of the case $\xi_1=0$). For example,
in the representation \eqref{dimmock_representation}:
\begin{displaymath}
 {\cal U}^\dag S(\xi_0,\xi_1)\,{\cal U}=\frac{2t}{\lambda}\xi_0+
 {\rm i}\sqrt{\xi_0^2+\xi_1^2}\,\beta{\bm\alpha}\cdot{\bm n}\equiv S_{\cal U}.
\end{displaymath}
The problem of determining the edge spectrum of the {\it original} Hamiltonian
$H$ [Eq. \eqref{dimmock_hamiltonian}] with the BC $S_{\cal U}$ is
mathematically identical to the analysis, which led to Eq.
\eqref{dimmock_ss_spectrum}. Of course, the transformation ${\cal U}$ changes
$H$ as well: $H\to H_{\cal U}\neq H$, but leaves its topological structure
intact. Qualitatively, variations in the boundary parameters $\xi_{0,1}$ can be
seen as unitary transformations of the bulk Hamiltonian. These transformations
also guarantee that the BC is kept diagonal. If initially the bulk band
structure was characterized by a strong topological index $\nu_0\neq0$, the
same will be true for $H_{\cal U}$. The shape of the edge spectrum does depend
on a particular choice of $\xi_{0,1}$ and has to be computed explicitly.
However, the above argument implies that if at $\xi_1=0$, the Fermi arc
enclosed an odd number (e.g. one) of Kramers-degenerate points, then the same
will be true for any $\xi_1$.

In Fig. \ref{fig_dimmock_sti_ss}, we present the dispersion relation of surface
states for several values of the BC parameters in the STI regime
($\epsilon=5.87t$) and for the trivial insulator ($\epsilon=6.37t$). From
panels (a)--(c) it follows that in the STI phase the number of edge states at
the Fermi level may change, depending on the BC parameters, but it {\it always}
remains odd, i.e. the parity of their number is protected. These surface states
exist for all values of $\xi_{0,1}$.

Panel (d) corresponds to the case $\epsilon>6t$ when the system is a trivial
insulator, which can nevertheless exhibit metallic edge states. The surface
spectrum is quite sensitive to the BC parameters and disappears for
$\xi_{0,1}\sim1$. Similar to the problem studied in the previous section, the
number of Fermi level crossings is even. This circumstance once again shows the
physical indistinguishability of a WTI and a trivial insulator.

\section{Effect of time-reversal breaking surface perturbations}
\label{sec:tri_break}
Time-reversal symmetry plays a crucial role in stabilizing the STI phase and
metallic properties of the surface. When this symmetry is broken via some
physical mechanism, the Kramers theorem no longer holds and surface Dirac
fermions are expected to acquire mass \cite{Kane_Z2,Zhang_MEE}, i.e. they
become gapped. In Ref. \onlinecite{Chen_2010} such gapped edge states were
indeed observed in ${\rm Bi_2Se_3}$ when doped with magnetic impurities at the
surface.

Phenomenologically, we can simulate this effect by adding a ${\cal T}$-breaking
perturbation to the boundary operator $S$. Let us consider one such term:
\begin{equation}
 \Delta S=\xi_{\cal T}\beta\gamma_5{\bm\alpha}\cdot{\bm n},
 \label{t_breaking_term}
\end{equation}
where $\gamma_5={\rm i}\alpha^x\alpha^y\alpha^z$ is the pseudo-scalar matrix
\cite{Akhiezer_Berestetskii}. Because of this matrix, the correction
\eqref{t_breaking_term} also breaks space inversion symmetry. Nevertheless, it
is easy to see that the current \eqref{dimmock_current} vanishes at the surface
after $\Delta S$ has been added to the boundary operator \eqref{dimmock_S}.

Since the operator $T_z$, introduced in Sec. \ref{sec:dirac_model}, is
invariant under time-reversal, $\Delta S$ will mix states with different values
of $\tau$. In the absence of a natural conserved quantity, suitable for
labelling single-electron states, the complete investigation of the edge
spectrum becomes quite cumbersome. Still, one can understand the qualitative
effect of $\Delta S$ by working in the perturbative regime, i.e. when
$\vert\xi_{\cal T}\vert$ is small. In order to simplify things even further, we
confine our analysis to a particular case, when there is no band mixing at the
surface, i.e. $\xi_1=0$ in \eqref{dimmock_S}. Then, we can use ideas of the
previous section to transfer the ${\cal T}$-breaking term from the BC to the
bulk Hamiltonian via a unitary transformation ${\cal U}_{\cal T}$, and employ
degenerate perturbation theory to treat the resulting corrections in the
Hamiltonian within the subspace of edge states \eqref{dimmock_ss_state}.

The unitary transformation ${\cal U}_{\cal T}$ is effected by $\gamma_5$ [cf.
Eq. \eqref{dimmock_S}]:
\begin{displaymath}
 {\cal U}_{\cal T}=\e^{{\rm i}\omega\gamma_5/2};\quad
 {\rm tg}\,\omega=\xi_{\cal T}/\xi_0.
\end{displaymath}
To first order in $\xi_{\cal T}$, the boundary operator \eqref{dimmock_S}
remains unchanged, and the only correction to the Hamiltonian
\eqref{dimmock_hamiltonian} is:
\begin{displaymath}
 \Delta H=\frac{\xi_{\cal T}}{\xi_0}\biggl[\epsilon\sum_i\Psid_i\Gamma\Psi_i-
 t\sum_{i,\mu}\bigl(\Psid_i\Gamma\Psi_{i+\mu}+{\rm h.c.}\bigr)\biggr]
\end{displaymath}
with $\Gamma={\rm i}\beta\gamma_5$. In the basis \eqref{dimmock_ss_state} the
matrix elements of $\Delta H$ can be written as:
\begin{displaymath}
 \langle\psi_{\tau^\prime\vkp}\vert\Delta H\vert\psi_{\tau\vkp}\rangle\sim
 \epsilon\bigl(\xi_{\cal T}/\xi_0\bigr)\sigma^x_{\tau^\prime\tau}.
\end{displaymath}
Therefore, at $\vkp=0$ (Dirac point) the perturbation \eqref{t_breaking_term}
opens a gap $\Delta_{\cal T}\sim\vert\xi_{\cal T}/\xi_0\vert$, in qualitative
agreement with the experiments of Ref. \onlinecite{Chen_2010}.

\section{Conclusion}
\label{sec:conclusion}
Since in real materials the electronic structure of the surface is largerly
unknown, investigation of the surface properties, such as the spectrum of
localized states, requires a phenomenological approach similar to the one
employed in the present work. It is important to realize that any bulk-boundary
correspondence cannot be simply based on general topological arguments that
disregard hypotheses about the boundary conditions used. The key idea of our
method consists of using the algebraic structure of the bulk Hamiltonian to
classify boundary conditions for the Bloch wavefunctions according to
fundamental symmetries of the problem.

When the surface preserves time-reversal symmetry, our results suggest that
experimentally, i.e. by looking at the edge spectrum, one can only discriminate
between the STI and trivial insulating phases. The claimed experimental
signature of a WTI -- an even number of surface states crossing the Fermi level
along a path between two TRIMs \cite{Kane_Z2} -- turns out to be misleading.
Indeed, by adjusting phenomenological parameters in the boundary conditions,
one can make a trivial insulator exhibit a surface spectrum similar to that of
a WTI, even in a clean system, where the periodicity of the surface is
preserved. Therefore, these two phases cannot be physically distinguished by
looking at the edge spectrum, and should be theoretically classified as the
same state.

On the contrary, in the STI phase the Fermi level crosses an odd number of
surface states. This precise number depends on a particular choice of the
boundary condition parameters, e.g. it changes from 1 to 3 in panels (a) and
(c) of Fig. \ref{fig_dimmock_sti_ss}, but its {\it parity} is always preserved.
Hence, the surface spectrum of a STI (and for the sake of the argument of any
band insulator) is not robust against time-reversal invariant boundary
perturbations.

These observations can be naturally summarized in a consistent definition of
a bulk-boundary correspondence:
\begin{displaymath}
 \nu_0=N_s\,\,{\rm mod}\,\,2,
\end{displaymath}
where $\nu_0$ is the strong $\z$ bulk topological invariant, introduced by Kane
{\it et al.} \cite{Kane_Z2} and $N_s$ is the number of surface Kramers doublets
inside the Fermi arc (or the number of edge states crossing the Fermi level
along any path between two TRIMs in the surface Brillouin zone). Only the
parity of the number of surface states is protected. This definition is
compatible with the classification of TIs according to quantization of the
axion $\theta$-term \cite{Zhang_MEE,Wang_2010}, which also describes the
orbital magneto-electric coupling \cite{Essin_2009}.

The bulk topological invariants are meaningful only if the surface preserves
time-reversal symmetry. In Sec. \ref{sec:tri_break} we demonstrated that a
${\cal T}$-breaking perturbation opens a gap in the edge spectrum at the TRIMs,
and effectively destroys the STI phase. This conclusion is in agreement with
recent experiments \cite{Chen_2010} performed in bismuth selenide with the
time-reversal symmetry explicitly broken by surface magnetic impurities.

Finally, we would like to emphasize that our demonstrations are applicable to
uncorrelated, i.e. mean-field-like, ``TI'' systems. The effect of correlations
beyond mean-field is still an open problem. While there are some efforts
\cite{Varney_2010} to extend the existing classification of TIs to interacting
systems, so far they only amount to mean-field arguments.

\end{document}